\documentclass{llncs}

\usepackage{amssymb}
\usepackage{pstricks,pst-plot} %
\usepackage{amsmath}%
\usepackage{hyperref}  
\hypersetup{
    colorlinks,%
    citecolor=black,%
    filecolor=black,%
    linkcolor=black,%
    urlcolor=black
}

 \title{The complexity of small universal {T}uring machines: a survey\thanks{This paper is extended and updated from~\cite{WoodsNeary2009A}. T.~Neary is support by Science Foundation Ireland, Grant Number 09/RFP/CMS2212. D.~Woods is supported by National Science Foundation Grant 0832824, the Molecular Programming Project. We thank Astrid Haberleitner for her tireless work in translating a number of highly technical papers from German to English, and  Beverley Henley for her support.}}

\author{Turlough Neary\inst{1} and Damien Woods\inst{2}}
\institute{School of Computer Science \& Informatics, University College Dublin, Ireland.\\ \email{turlough.neary@ucd.ie} \and Division of Engineering \& Applied Science, California Institute of Technology, Pasadena, CA 91125, USA. \\  \email{woods@caltech.edu}}
\begin{document}
\maketitle

\begin{abstract}
We survey some work concerned with small universal Turing machines, cellular automata, tag systems, and other simple models of computation. For example it has been an open question for some time as to whether the smallest known universal Turing machines of Minsky, Rogozhin, Baiocchi and Kudlek are efficient (polynomial time) simulators of Turing machines. These are some of the most intuitively simple computational devices and previously the best known simulations were exponentially slow. We discuss recent work that shows that these machines are indeed efficient simulators. In addition, another related result shows that Rule~110, a well-known elementary cellular automaton, is efficiently universal. 
We also discuss some old and new universal program size results, including the smallest known universal Turing machines. We finish the survey with results on generalised and restricted Turing machine models including machines with a periodic background on the tape (instead of a blank symbol), multiple tapes, multiple dimensions, and machines that never write to their tape. 
We then discuss some ideas for future work.
\end{abstract}

\section{Introduction}\label{sec:Introduction}
In this survey we explore results related to the time and program size complexity of universal Turing machines, and other models of computation. We also discuss results for variants on the Turing machine model to give an idea of the many strands of work in the area. Of course the choice of topics is incomplete and reflects the authors' interests, and there are other related surveys that may interest the reader~\cite{Mar2006p-acc,Margenstern1997A,Mar2000p,MooreMertens2011}.

\begin{figure}[t]
\begin{center}
\newcommand{\dwtnfigurefontsize}{\scriptsize}
\newcommand{\dwtnfigurelegendfontsize}{\scriptsize}

\psset{unit=2.4ex}
\begin{pspicture}(-3,-2.5)(29,19) %
\put (7.4,17) {\dwtnfigurelegendfontsize : universal, direct simulation, $O(t^2)$,~\cite{NW2006p}}

\psset{dotsize=5pt,dotstyle=o}
\psdot (6.8,16.2)
\put (7.4,16) {\dwtnfigurelegendfontsize : universal, 2-tag simulation, $O(t^4 \log^2 t)$,~\cite{Bai2001c,KR2001c,Rog1996p}}

\psset{dotsize=5.5pt,dotstyle=triangle*}
\psdot (6.8,15.2)
\put (7.4,15) {\dwtnfigurelegendfontsize : universal, bi-tag simulation, $O(t^6)$,~\cite{NearyWoods2009}}

\psset{dotstyle=diamond,linewidth=1pt}
\psdot (6.8,14.2)
\put (7.4,14) {\dwtnfigurelegendfontsize : semi-weakly universal, direct simulation, $O(t^2)$,~\cite{Wat1972p}}

\psset{dotstyle=diamond*,dotsize=4.75pt}
\psdot (6.8,13.2)
\put (7.4,13) {\dwtnfigurelegendfontsize : semi-weakly universal, cyclic-tag simulation, $O(t^4 \log^2 t)$,~\cite{WoodsNeary2009}}

\psset{dotstyle=square*,dotsize=5pt}
\psdot (6.8,12.2)
\put (7.4,12) {\dwtnfigurelegendfontsize : weakly universal, Rule 110 simulation, $O(t^4 \log^2 t)$,~\cite{NearyWoods2009A}}

\psset{dotstyle=square,dotsize=5pt}
\psdot (6.8,17.2)
\put (20.2,1.8) {\dwtnfigurelegendfontsize universal}

\put (20.2,0.7) {\dwtnfigurelegendfontsize non-universal}

\psset{linestyle=solid}
{\tiny \psaxes[ticksize=1pt]{->}(20,19) }
\rput (4.5,-2) {\dwtnfigurefontsize states}
\put (-5,4) {\dwtnfigurefontsize symbols}

\psline (3,1)(20,1)
\psline (1,3)(1,19)
\psline (2,2)(3,2)
\psline (2,2)(2,3)
\psline (1,3)(2,3)
\psline (3,1)(3,2)

\psset{linestyle=dashed, dash=3pt 3pt}
\psline (2,18)(2,19) 
\psline (2,18)(3,18)
\psline (3,9)(3,18)
\psline (3,9)(4,9)
\psline (4,6)(4,9)
\psline (4,6)(5,6)
\psline (5,5)(5,6)
\psline (5,5)(6,5)
\psline (6,4)(6,5)
\psline (6,4)(9,4)
\psline (9,3)(9,4)
\psline (9,3)(15,3)
\psline (15,2)(15,3)
\psline (15,2)(20,2)

\psset{dotsize=5.5pt}
\psset{dotstyle=triangle*}
\psdot (9,3)
\psdot (15,2)
\psdot (6,4)
\psset{dotsize=7pt}
\psdot (5,5)

\psset{dotsize=5pt,dotstyle=o}
\psdot (2,18)
\psdot (3,9)
\psdot (4,6)
{\psset{dotsize=4.5pt} \psdot (5,5)}
\psdot (7,4)
\psdot (10,3)
\psdot (19,2)

\psset{dotstyle=square,dotsize=5.5pt}
\psdot (3,11)
\psdot (5,7)
\psdot (6,6)
\psdot (7,5)
\psdot (8,4)

\psset{dotstyle=square*,dotsize=5pt}
\psdot (2,4)
\psdot (3,3)
\psdot (6,2)

\psset{dotsize=4.75pt,linewidth=1pt} 
\psset{dotstyle=diamond,linewidth=1pt}
\psdot (5,4)
\psdot (7,3)

\psset{dotsize=4.75pt}
\psset{dotstyle=diamond*}
\psdot (4,5)
\psdot (3,7)
\psdot (2,13)

\end{pspicture}
\end{center}
\caption{State-symbol plot of small universal Turing machines.  
The type of simulation is given for each group of machines. 
Simulation time overheads are given in terms of simulating a single-tape deterministic Turing machine that runs in time $t$.}
\label{fig:states-sym-jan07}
\end{figure}
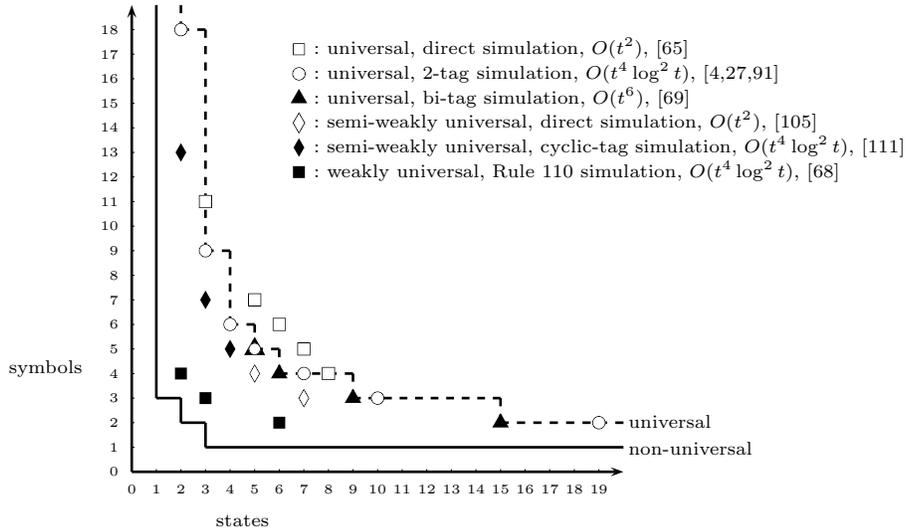

In 1956 Shannon~\cite{Sha1956p} considered the question of finding the smallest possible universal Turing machine~\cite{Turing1937A}, where size is the number of states and symbols.  In the early Sixties, Minsky and Watanabe had a running competition to see who could find the smallest universal Turing machine~\cite{Min1960m,Min1962p,Wat1960m,Wat1961p}. Early attempts~\cite{Ike1958x,Wat1961p} gave small universal Turing machines that efficiently (in polynomial time) simulated Turing machines.  In 1962, Minsky~\cite{Min1962p} found a small 7-state, 4-symbol universal machine. Minsky's machine worked by simulating 2-tag systems, which were shown to be universal by Cocke and Minsky~\cite{CM1964p,Min1962m}.  Rogozhin~\cite{Rog1982p} extended Minsky's technique of 2-tag simulation and found small machines with a number of state-symbol pairs. Subsequently, some of Rogozhin's machines were reduced in size or improved by Robinson~\cite{Rob1991p,Rog1996p}, Kudlek and Rogozhin~\cite{KR2001c}, and Baiocchi~\cite{Bai2001c}. All of the smallest known 2-tag simulators are plotted as circles in Figure~\ref{fig:states-sym-jan07}. Also, Table~\ref{tab:standard} lists a number of these machines.

Unfortunately, Cocke and Minsky's 2-tag simulation of Turing machines was exponentially slow. The exponential slowdown was essentially caused by the use of a unary encoding of Turing machine tape contents. Therefore, for many years it was entirely plausible that there was an exponential trade-off between program size complexity on the one hand, and time/space complexity on the other: the smallest universal Turing machines seemed to be exponentially slow. 

Figure~\ref{fig:states-sym-jan07} shows a non-universal curve. This curve is a lower bound that gives the state-symbol pairs for which it is known that the halting problem is decidable. The 1-symbol case is trivial and Shannon~\cite{Sha1956p} claimed that 1-state Turing machines are non-universal. However, both Fischer~\cite{Fischer1965} and Nozaki~\cite{Nozaki1969} noted that Shannon's definition of universal Turing machine is too strict and so his proof is not sufficiently general. Later, the 1-state case was shown by Hermann~\cite{Her1968c}.
Pavlotskaya~\cite{Pav1973p} and, via another method, Kudlek~\cite{Kud1996p} have shown that there are no universal 2-state, 2-symbol machines, where one transition rule is reserved for halting. Pavlotskaya~\cite{Pav1978p} has also shown that there are no universal 3-state, 2-symbol  machines, and also claimed~\cite{Pav1973p}, without publishing a proof, that there are no universal machines for the 2-state, 3-symbol case. Again, both of these cases assume that a transition rule is reserved for halting.

\section{Time and size efficiency of universal machines}
As mentioned above, some of the very earliest small Turing machines were polynomial time simulators. Subsequently, attention turned to the smaller, but exponentially slower, 2-tag simulators given by Minsky, Rogozhin and others.

\begin{table}[t]
\begin{center}
\begin{tabular}{@{}c@{\;\:}@{\;\,}c@{\;\;\;\,}c@{\;\;}l@{}}
 states & symbols & state-symbol product & \;\;author  \\ \hline
$m$ & 2 & 2$m$  & Shannon~\cite{Sha1956p}  \\
2 & $n$ &  2$n$ & Shannon~\cite{Sha1956p}  \\
12 & 6 &  72 &  Takahashi~\cite{Takahashi1958} (mentioned in~\cite{Wat1961p}) \\
10 & 6 & 60   &  Ikeno~\cite{Ike1958x} (also appears in~\cite{Min1960m})  \\
8 & 6 &  48 &  Watanabe~\cite{Wat1960m} (mentioned in~\cite{Min1962p}) \\
7 & 6 &  42 &  Minsky~\cite{Min1960m} \\
8 & 5 & 40 & Watanabe~\cite{Wat1961p} \\
9 & 4 & 36 &  Tritter (mentioned in~\cite{Min1962p}) \\
25 & 2 & 50 &  Minsky~\cite{Min1962m} \\
6 & 6 & 36 &  Minsky~\cite{Min1962p} \\
7 & 4 & 28 &  Minsky~\cite{Min1962p} \\
24 & 2 & 48 &  Rogozhin~\cite{Rogozhin1979,Rog1982p,Rog1996p}  \\
2 & 21 & 42 &  Rogozhin~\cite{Rogozhin1979,Rog1982p}  \\
11 & 3 & 33 &  Rogozhin~\cite{Rogozhin1979,Rog1982p} \\
3 & 10 & 30 &  Rogozhin~\cite{Rogozhin1979,Rog1982p} \\
7 & 4 & 28 &  Rogozhin~\cite{Rogozhin1979,Rog1982p,Rog1996p} \\
5 & 5 & 25 &  Rogozhin~\cite{Rogozhin1979,Rog1982p,Rog1996p} \\
4 & 6 & 24 &  Rogozhin~\cite{Rogozhin1979,Rog1982p,Rog1996p} \\
2 & 18 & 36 &  Rogozhin~\cite{Rog1996p} \\
10 & 3 & 30 &  Rogozhin~\cite{Rogozhin1992,Rog1996p} \\
3 & 10 & 30 &  Rogozhin~\cite{Rogozhin1993,Rog1996p}*  \\
22 & 2 & 44 &  Rogozhin~\cite{Rogozhin1998} \\
19 & 2 & 38 &  Baiocchi~\cite{Bai2001c} \\
7 & 4 & 28 &  Baiocchi~\cite{Bai2001c}* \\
3 & 9 & 27 &  Kudlek \& Rogozhin~\cite{KR2001c} \\
18 & 2 & 36 &  Neary \& Woods~\cite{NW2007c} \\
9 & 3 & 27 &  Neary \& Woods~\cite{NearyWoods2009} \\
5 & 5 & 25 &  Neary \& Woods~\cite{NearyWoods2009}* \\
6 & 4 & 24 &  Neary \& Woods~\cite{NearyWoods2009} \\
15 & 2 & 30 &  Neary \& Woods~\cite{NearyWoods2009} \\
\hline  
\end{tabular}
\end{center}
\caption{Small standard universal Turing machines, ordered by date and then by state-symbol product. If there are multiple machines with the same state-symbol pair, the machine with the smallest number of instructions is denoted~*.}\label{tab:standard}
\end{table}

Recently~\cite{NW2006p} we have given small machines that are efficient polynomial time simulators. More precisely, if~$M$ is a deterministic single-tape Turing machine that runs in time $t$ and space $s$, then there are machines, with state-symbol pairs given by the squares in Figure~\ref{fig:states-sym-jan07}, that directly simulate~$M$ in polynomial time~$O(t^{2})$ and linear space $O(s)$. These machines define a~$O(t^2)$ curve.  They are currently the smallest known universal Turing machines that simulate Turing machines in~$O(t^2)$ time. Their $O(s)$ space usage is also extremely efficient, more efficient than the other machines in Figure~\ref{fig:states-sym-jan07}, all of which use space that is up to square root of their simulation time.

Despite the existence of these efficient $O(t^2)$ simulators, it still remained the case that the 
smallest universal machines were exponentially slow. However, we have recently shown that the 
smallest machines are in fact efficient simulators of Turing machines, by showing that 2-tag systems 
are efficient~\cite{WN2006c}.  Tag systems are one of a number of rewriting systems invented in the 
1920s by Post, although published somewhat later~\cite{Post1943}. Post wanted to prove the 
decidability of various properties of tag systems, but found that even very simple examples had 
extremely complicated behaviour. Forty years later, Minsky showed that tag systems~\cite{Minsky1961} 
are in fact computationally universal, and then Cocke and Minsky~\cite{CM1964p,Min1962m} showed 
universality for a particularly simple form called 2-tag systems. Minsky~\cite{Min1962p,Min1962m} 
saw that one could find very small universal Turing machines by simulating 2-tag systems, and since 
then 2-tag systems have been at the core of many results in the field.

A 2-tag system acts on a dataword, which is a string of symbols taken from a finite 
alphabet~$\Sigma$. There is a fixed set of rules $R: \Sigma \rightarrow \Sigma^*$. In a single 
timestep, the leftmost symbol $\sigma_j$ of the dataword is read,  if there is a rule $\sigma_j 
\rightarrow \alpha_j$ then the string $\alpha_j$ is appended to the right of the dataword and the 
leftmost $two$ dataword symbols are deleted. This process is iterated until a suitable halting 
condition is reached (i.e.~there is no rule for the read symbol, the dataword has length less than 
$2$, or the 2-tag system enters a repeating loop). Part of the reason why it was presumed that 2-tag 
systems were exponentially slow is that it is not obvious how to locate a specific symbol based 
solely on its position relative to other symbols in the dataword (one might want to do this to 
simulate the local action of a Turing machine tape head). The main result of~\cite{WN2006c} uses an 
algorithm that solves this problem, and does so efficiently.

More precisely, given a deterministic single-tape Turing machine $M$ that runs in time $t$, there is a 2-tag system that simulates $M$ and runs in polynomial time $O(t^{4}\log^{2} t)$. The small machines of Minsky, Rogozhin, and others have a quadratic time  overhead when simulating 2-tag systems, hence by the result in~\cite{WN2006c} they simulate Turing machines in time $O(t^{8}\log^{4} t)$. It turns out that the time overhead can be improved~\cite{Nea2008m} to $O(t^{4}\log^{2} t)$, giving the $O(t^{4}\log^{2} t)$ time overhead for the machines shown in Figure~\ref{fig:states-sym-jan07} as hollow circles. Thus, there is currently little evidence for the claim of an exponential trade-off between program size complexity, and time/space complexity.

From the point of view of program size, Neary and Woods~\cite{Nea2008m,NearyWoods2009} have recently given four Turing machines that are presently the smallest known (standard) machines with 2, 3, 4 and 5 symbols. The 5-symbol machine improves on the 5-symbol machine of Rogozhin~\cite{Rog1996p} by one transition rule. The remainder of these machines improve on the 2- and 4-symbol machines of Baiocchi~\cite{Bai2001c}, and the 3-symbol machine of Rogozhin~\cite{Rog1996p}. These small machines simulate Turing machines in polynomial time~$O(t^6)$ and are illustrated as triangles in Figure~\ref{fig:states-sym-jan07}. They were proven universal via simulation of our universal variant of tag systems called {\em bi-tag systems}~\cite{NearyWoods2009}. Bi-tag systems are essentially 1-tag systems (and so they read and delete one symbol per timestep) augmented with additional context sensitive rules that read, and delete, two symbols per timestep. Bi-tag systems are a restriction of Post's normal systems~\cite{Post1943}. On the one hand bi-tag systems are universal, while on the other hand they are sufficiently `simple' to be simulated by such small machines.

Exponentially improving the time efficiency of 2-tag systems has implications for a number of models of computation, besides small universal Turing machines.  Following our result, the simulation efficiency of many biologically inspired models of computation, including neural networks, H systems and P systems, has been improved from exponential to polynomial. 
For example, Siegelmann and Margenstern~\cite{SM1999p} give a neural network that uses only nine high-order neurons to simulate 2-tag systems.  Taking each synchronous update of the nine neurons as a single parallel timestep, their neural network simulates 2-tag systems in linear time.  They note that ``tag systems suffer a significant slow-down ... and thus our result proves only Turing universality and should not be interpreted complexity-wise as a Turing equivalent.''  Now we know that their neural network is in fact efficiently universal.
Rogozhin and Verlan~\cite{RV2005c} give a tissue P system with eight rules that simulates 2-tag systems in linear time, and thus we have improved its simulation time overhead from exponential to polynomial.  This system uses splicing rules (from H systems) with membranes (from P systems) and is non-deterministic.   Harju and Margenstern~\cite{HM2005} gave an extended H-system with 280 rules that generates recursively enumerable sets using Rogozhin's 7-state, 4-symbol universal Turing machine. Using our result from 2-tag systems, the time efficiency of their construction is improved from exponential to polynomial, with a possible small constant increase in the number of rules. The efficiency of Hooper's~\cite{Hooper1969} small 2-tape universal Turing machine is also improved from exponential to polynomial, as is Rothemund's~\cite{Rothemund1996} restriction enzyme implementation of Minsky's 7-state, 4-symbol UTM. The technique of simulation via 2-tag systems is at the core of many of the universality proofs in Margenstern's survey~\cite{Mar2000p}. Our work exponentially improves the time overheads in these simulations, such as Lindgren and Nordahl's cellular automata~\cite{LN1990p}, Margenstern's non-erasing Turing machines~\cite{Mar1993c,Mar1995c}, and Robinson's tiling~\cite{Rob1971p}.

\section{Non-standard universal Turing machines: time efficiency and program size}
So far we have been discussing results for universal Turing machines that have one tape, one tape head, and are deterministic (we often refer to this setup as the {\em standard} model). Of course one can consider results for other variants of the model. There are many generalised models, for example allowing multiple tapes, multiple dimensions, or even coupling the Turing machine with a finite automaton. 
Restricted models include non-printing, non-erasing and reversible Turing machines, and machines with restricted instructions. In this section we explore program size and time complexity results for a number of generalised and restricted models. Table~\ref{tab:non-standard} contains program size results for a number of such non-standard machines.

\begin{table}
\begin{center}
\begin{tabular}{@{\;}c@{\;\;\,}c@{\;\;\;}c@{\;\;\;}c@{\;\;}c@{\;\,}}
states & symbols & dimensions & tape & author  \\ \hline
15 & 2 & 1 & 3 & Moore~\cite{Moore1952}$\dagger$ \\
6 & 5 & 1 & 1 & Watanabe~\cite{Wat1961p}$\dagger$\\
1 & 2 & 1 & 4 & Hooper~\cite{Hooper1963,Hooper1969}$\dagger$ \\
2 & 3 & 1 & 2 & Hooper~\cite{Hooper1963,Hooper1969} \\
7 & 3 & 1 & 1 & Watanabe (mentioned in~\cite{Wat1972p,Nozaki1969})$\dagger$\\
5 & 4 & 1 & 1 & Watanabe~\cite{Wat1972p}$\dagger$\\
8 & 4 & 2 & 1 & Wagner~\cite{Wagner1973} \\
2 & 7 & 2 & 1 & Ottmann~\cite{Ottmann1975A}$\ddagger$ \\
10 & 2 & 2 & 1 & Ottmann~\cite{Ottmann1975B,KleineOttmann1977}$\ddagger$ \\
6 & 3 & 2 & 1 & Ottmann~\cite{Ottmann1975B,KleineOttmann1977}$\ddagger$ \\
4 & 4 & 2 & 1 & Ottmann~\cite{Ottmann1975B,KleineOttmann1977}$\ddagger$ \\
2 & 6 & 2 & 1 & Kleine-B\"uning \& Ottmann~\cite{KleineOttmann1977}$\ddagger$\\
2 & 5 & 2 & 1 & Kleine-B\"uning \& Ottmann~\cite{KleineOttmann1977}$\ddagger$\\
2 & 3 & 2 & 1 & Kleine-B\"uning \& Ottmann~\cite{KleineOttmann1977}$\ddagger$ \\
1 & 7 & 3 & 1 & Kleine-B\"uning \& Ottmann~\cite{KleineOttmann1977}$\ddagger$ \\
4 & 5 & 2 & 1 & Kleine-B\"uning \& Ottmann~\cite{KleineOttmann1977}\\
3 & 6 & 2 & 1 & Kleine-B\"uning \& Ottmann~\cite{KleineOttmann1977} \\
10 & 2 & 2 & 1 & Kleine-B\"uning~\cite{Kleine1977}\\
2 & 5 & 2 & 1 & Kleine-B\"uning~\cite{Kleine1977}\\
2 & 4 & 2 & 1 & Priese~\cite{Pri1979p} \\
4 & 2 & 2 & 1 & Gajardo et al.~\cite{Gajardo2002} \\
2 & 2 & 2 & 1 & Priese~\cite{Pri1979p}$\vartriangle$ \\
2 & 5 & 1 & 1 & Margenstern \& Pavlotskaya~\cite{MargensternPavlotskaya1995A}$\star$\\
4 & 7 & 1 & 1 & Pavlotskaya~\cite{Pavlotskaya1996}$\star$\\
2 & 3 & 1 & 1 & Margenstern \& Pavlotskaya~\cite{MP2003p}$\star$\\
7 & 2 & 1 & 1 & Eppstein (published by Cook~\cite{Coo2004p})$\ddagger$\\
4 & 3 & 1 & 1 & Cook~\cite{Coo2004p} \& Wolfram~\cite{Wol2002x}$\ddagger$\\
3 & 4 & 1 & 1 & Cook~\cite{Coo2004p} \& Wolfram~\cite{Wol2002x}$\ddagger$\\
2 & 5 & 1 & 1 & Cook~\cite{Coo2004p} \& Wolfram~\cite{Wol2002x}$\ddagger$\\
6 & 2 & 1 & 1 & Neary \& Woods~\cite{NearyWoods2009A}$\ddagger$\\
3 & 3 & 1 & 1 & Neary \& Woods~\cite{NearyWoods2009A}$\ddagger$\\
2 & 4 & 1 & 1 & Neary \& Woods~\cite{NearyWoods2009A}$\ddagger$\\
3 & 7 & 1 & 1 & Woods \& Neary~\cite{WoodsNeary2009}$\dagger$\\
4 & 5 & 1 & 1 & Woods \& Neary~\cite{WoodsNeary2009}$\dagger$\\
2 & 13 & 1 & 1 & Woods \& Neary~\cite{WoodsNeary2009}$\dagger$\\
\hline
\end{tabular}
\end{center}
\caption{Small non-standard universal Turing machines. Semi-weak machines are denoted by~$\dagger$, weak machines by~$\ddagger$, machines coupled with a finite automaton by $\star$, and a machine with 2 tape heads by~$\vartriangle$.}\label{tab:non-standard}
\end{table}

\subsection{Weak universality and Rule~110}
An interesting generalisation occurs when we stick to the standard conventions, but we allow the blank portion of the tape to contain a word, that is constant (independent of the input), and is repeated infinitely often in one direction, say to the left of the input. We say that such Turing machines are {\em semi-weakly universal}. Some of the earliest small universal Turing machines were semi-weak~\cite{Wat1961p,Wat1972p}. Sometimes another word is also repeated infinitely often to the right. Universal machines that use this setup are called {\em weakly universal}~\cite{Mar2006c}. 

It is not difficult to see how this generalisation can help to reduce program size. For example, it is typical of small universal Turing machine simulations that the program being simulated is stored on the tape. When reading an instruction we often mark certain symbols. At a later time we then restore marked symbols to their original values. If the simulated program is repeated infinitely often, say to the left of the input, things may be much easier as we can simply skip the `restore' phase of our algorithm and access a new copy of the program when simulating the next instruction, thus reducing the universal program's size.

This was the strategy used by Watanabe~\cite{Wat1961p,Wat1972p} to find the semi-weak, direct Turing machine simulators shown as hollow diamonds in Figure~\ref{fig:states-sym-jan07}.
Recently~\cite{WoodsNeary2009} we have given three new semi-weakly universal machines and these are shown as solid diamonds in Figure~\ref{fig:states-sym-jan07}. These machines simulate cyclic tag systems~\cite{Coo2004p}.
It is interesting to note that two of our machines are symmetric with those of Watanabe (around the line where states $=$ symbols), despite the fact that we use a different simulation technique. Our 4-state, 5-symbol machine has only 17 transition rules, making it the smallest known semi-weakly universal machine (Watanabe's 5-state, 4-symbol machine has 18 transition rules, and his 7-state, 3-symbol machine has 21 rules~\cite{Wat1972p})\footnote{Watanabe mentions that he found a $(7,3)$ universal machine with 21 transition rules in reference~\cite{Wat1972p}. We have not found the details of this machine, however the most reasonable inference from the literature is that it is semi-weakly universal.}. The time overhead for these machines is polynomial. More precisely, if $M$ is a single-tape deterministic Turing machine that runs in time $t$, then $M$ is simulated by either of our semi-weak machines in time $O(t^4 \log^2 t)$. Watanabe's semi-weak machines also ran in polynomial time, with a very efficient time overhead of $O(t^2)$.

Cook, Eppstein, and Wolfram~\cite{Coo2004p,Wol2002x} gave weakly universal Turing machines that were significantly smaller than the existing semi-weak machines. These were improved upon by Neary and Woods~\cite{NearyWoods2009A} to give the smallest known weakly universal machines. In (states, symbols) notation their sizes are $(2,4)$, $(3,3)$ and $(6,2)$, and they are illustrated in Figure~\ref{fig:states-sym-jan07}. These machines work by simulating Rule~110, a very simple kind of cellular automaton. Rule~110 is an elementary cellular automaton, which means that it is a one-dimensional, nearest neighbour, binary cellular 
automaton~\cite{Wol1983p}. More precisely, it is composed of a sequence of cells $\ldots 
p_{-1}p_{0}p_{1}\ldots$ where each cell has a binary state $p_i\in\{0,1\}$.  
At timestep $t+1$ the value of cell 
$p_{i,t+1}=F(p_{i-1,t},p_{i,t},p_{i+1,t})$ is given by the synchronous local 
update function $F$
\begin{align*}\label{eq:Rule110}
\begin{split}
F(0,0,0)=0 \qquad & \qquad  F(1,0,0)=0 \\
F(0,0,1)=1 \qquad & \qquad F(1,0,1)=1 \\
F(0,1,0)=1 \qquad & \qquad F(1,1,0)=1 \\
F(0,1,1)=1 \qquad & \qquad F(1,1,1)=0
\end{split}
\end{align*}
Rule~110 was shown to be universal via an impressive and detailed simulation of cyclic tag systems, the result is stated and described in~\cite{Wol2002x} and the full proof is given in~\cite{Coo2004p}. In the proof, the Rule~110 instance has a special (constant) word repeated infinitely to the left of the input, and another to the right. Rule~110 has a very simple update rule which facilitates the writing of very small weak Turing machines to simulate it.

As noted, Rule~110 was shown to be universal by simulating cyclic tag systems, which in turn simulate 2-tag systems. The chain of simulations included the exponentially slow 2-tag algorithm of Cocke and Minsky, thus Rule~110, and the weakly universal machines that simulate it, were exponentially slow. In a recent paper~\cite{NW2006c} we have improved their simulation time overhead to polynomial by showing that cyclic tag systems are efficient simulators of Turing machines. In doing so, we solved what Cook~\cite{Cook2009} has called the ``geometry problem of cyclic-state tape processors." The difficult in overcoming this problem is that there is no obvious way for the system to efficiently determine which symbols or objects are adjacent to each other. Previous works used unary encodings as it was not obvious how to determine the relative positions of adjacent digits in a sequence. Our main result was in providing an efficient solution to this problem.

Our result has interesting implications for Rule~110. For example, given an initial configuration of Rule~110, and a value~$t$ in unary, predicting~$t$ timesteps of a Rule~110 computation is P-complete. Therefore, unless $\mathrm{P}=\mathrm{NC}$, which is widely believed to be false, we cannot hope to quickly (in polylogarithmic time) predict the evolution of this simple cellular automaton even if we have a polynomial amount of parallel hardware. Rule~110 is the simplest (one-dimensional, nearest neighbour) cellular automaton that has been shown to have a P-complete prediction problem. In particular, Ollinger's~\cite{Oll2002c} intrinsic universality result already shows that prediction for one dimensional nearest neighbour cellular automata is P-complete for~6 states (later improved to~4 states by Richard and Ollinger~\cite{Ric2006m,OllingerRichard2011}), and our result improves this to 2 states.  The question of whether Rule~110 prediction is P-complete has been asked, directly or indirectly, in a number of previous works (for  example~\cite{Aar2002p,Moo1999p,Moo1999bp}).	

It is currently unknown whether all of the lower bounds in Figure~\ref{fig:states-sym-jan07} hold for weak machines. For example, the non-universality results of Pavlotskaya were proven for the case where one transition rule is reserved for halting, however the smallest weak machines do not halt.

\subsection{Other non-standard universal Turing machines}\label{sec:nonStandard-UTMs}
Weakness has not been the only generalisation on the standard model in the search for ever smaller universal machines. We give some notable examples here, many others are to be found in Table~\ref{tab:non-standard}. 

Before Shannon's famous paper, Moore~\cite{Moore1952} observed that 2-symbol machines were universal as any Turing machine could be converted into a 2-symbol machine by the (now) usual encoding. In the same paper Moore used this observation to give a universal 3-tape machine with 15 states and 2 symbols. Moore's machine uses only 57 instructions, each instruction being a sextuple that either moves one of its tape heads or prints a single symbol to one of its tapes. One of the tapes in Moore's 3-tape machine is circular and contains the simulated program, therefore his machine also operates correctly if the circular tape is replaced with a one-way infinite tape with a periodic background (i.e.~semi-weak). Moore's result has been largely ignored in the literature despite being the first published small universal Turing machine. Interestingly, Moore's paper cites unpublished work by Shannon on the universality of non-erasing machines.

Hooper~\cite{Hooper1963,Hooper1969} gave universal machines with 2 states, 3 symbols and 2 tapes, and with 1 state, 2 symbols and 4 tapes. One of the tapes in Hooper's 4-tape machine is circular and contains the simulated program, and so could be replaced by a one-way infinite tape with a periodic background (i.e.~semi-weak). Priese~\cite{Pri1979p} gave a 2-state, 4-symbol machine with a 2-dimensional tape, and a 2-state, 2-symbol machine with 2 tape heads and a 2-dimensional tape. Margenstern and Pavlotskaya~\cite{MP2003p,MargensternPavlotskaya1995A} gave a 2-state, 3-symbol Turing machine that uses only 5 instructions and is universal when coupled with a finite automaton. They also showed that the halting problem is decidable for such machines with 4 instructions~\cite{MP2003p}.

\subsection{Restricted universal Turing machines}
If we suitably restrict the standard Turing machine model the problem of finding universal machines with small state-symbol products 
becomes more difficult. Over the years, a number of authors have looked at non-erasing Turing machines, that is machines that
are permitted to overwrite blank symbols only. Moore~\cite{Moore1952} mentions that Shannon had proved that such non-erasing Turing machines simulate arbitrary Turing machines, however Shannon's work was never published. Shortly after, Shannon published a proof that 2-symbol Turing machines are universal, and Wang~\cite{Wang1957} proved that 2-symbol non-erasing Turing machines are universal. 
Later, Minsky proved the same result as Wang, but using the technique of simulation via non-writing Turing machines, yet another (universal) restriction~\cite{Minsky1961}.

Margenstern has examined the universality of 2-symbol Turing machines for a number of different restrictions. 
One such restriction is the number of colours of a machine, defined as the number of distinct triples $(\alpha,D,\delta)$, where $\alpha$ is the read symbol, $D$ is the move direction, and $\delta$ is the write symbol of a transition rule. 
Pavlotskaya~\cite{Pav1973p,Pavlotskaya1975} has shown that there are standard universal Turing machines with 3 colours
and no standard universal Turing machines with 2 colours. 
Margenstern~\cite{Mar1993c} has shown that there are non-erasing universal Turing machines with 5 colours and no non-erasing universal Turing machines with 4 colours.  
Laterality number is another property examined by Margenstern. 
The laterality number of a Turing machine is defined as the minimum of the number of left move instructions and the number of right move instructions. 
Margenstern and Pavlotskaya~\cite{Pav1973p,MargensternPavlotskaya1995B} have shown that there are universal Turing machines with laterality number~2 and no universal machines with laterality number 1. 
Margenstern~\cite{Mar1995c,Mar1997p} has shown that there are universal non-erasing Turing machines with laterality number~3 and no universal non-erasing machines with laterality number 2. 
For more on these results see~\cite{Margenstern1992,Mar1993c,Margenstern1994,Mar1995c,Margenstern1995B,Margenstern2001}.

Fischer~\cite{Fischer1965} gives a number of universality results for Turing machines that use restricted forms of transition rules. In one result he proves that 3-state Post machines are universal (Post machines~\cite{Pos1947p} are like Turing machines, except that in a single timestep they can move or write, but not both). Interestingly, Aanderaa and Fischer~\cite{AF1967p} show that the halting problem for 2-state Post machines is decidable.

Bennett~\cite{Bennett1973} has shown that 3-tape reversible Turing machines are universal. Morita and others have since shown universality results for reversible Turing machines with 1 tape and 2 symbols~\cite{Morita1989}, and 17 states and 5 symbols~\cite{MY2007c}.

\subsection{Universal Turing machines with multidimensional tapes: time efficiency and program size}
During the 1970s a number of authors~\cite{Pri1979p,KleineOttmann1977,Wagner1973} were interested in finding small universal Turing machines with multidimensional tapes. 
The machines of these authors have not, to our knowledge, been analysed from the perspective of time/space complexity. We discuss this topic here.

Lutz Priese~\cite{Pri1979p} gives a 2D machine with 2 states and 4 symbols that is universal on finite initial conditions (i.e.\ all except a finite number of symbols are initially blank), and another 2D machine with 2 states, 2 symbols and 2 tape heads that is derived from this 4-symbol machine.
Priese's machines simulate counter machines (also called register machines), via a sequence of reductions. 
Given a counter machine that runs in time $\tau$, Priese's machines simulate its computation in time $O(\tau^2)$ and space $O(\tau)$. 
Due to the unary encoding used by counter machines~\cite{Fischer1968}, both of Priese's machines simulate Turing machines with an exponential time overhead. 
Priese's machines do not end their computation in the conventional manner of halting on a state-symbol pair that has no transition rule: 
instead there is a choice, via the initial input encoding, of ending a computation either by entering a sequence of 6 repeating configurations or by halting when an attempt is made to move off the edge of the 2D tape. 

Langton's ant~\cite{Langton1986} is usually described as an ant that lives on a 2D grid of binary-valued cells. The ant chooses which adjacent cell to move to based on (a)~the current cell's binary value and (b) the ant's current orientation. The ant flips the current cell's bit as it moves away. So Langton's ant is a 2D Turing machine with 2 symbols and 4 states (North, South, East and West). Gajardo et al.~\cite{Gajardo2002} showed that predicting the behaviour of the ant is P-hard, by simulating Boolean circuits in polynomial time. By then showing how the ant can simulate an infinite sized circuit (with a simple repeating structure), which in turn can simulate the space-time diagram of a cellular automata (CA), they prove that Langton's ant is weakly universal in 2D.

It is worth pausing to describe a form of weak universality in 2D, where the tape has a background that is ultimately periodic in both dimensions of single quadrant. A one-way infinite sequence is ultimately periodic~\cite{Friedman1962} if it is of the form $s_1 s_2^{\omega}$ where $s_2^{\omega}=s^2s^2s^2\ldots$, and $s_1$ and $s_2$ are finite sequences.
We say that a $\mathbb{N}\times\mathbb{N}$ pattern is ultimately periodic in the $x$ direction if for each $y \in\mathbb{N}$ the infinite sequence of symbols at the coordinates $(0,y),(1,y),(2,y),\ldots\,$ is ultimately periodic. This is defined analogously for the $y$ direction.

Kleine-B\"uning and Ottmann~\cite{KleineOttmann1977} give universal Turing machines which have a single multidimensional tape, a number of which are weakly universal. 
Remarkably, their 2D, 2-state, 3-symbol machine does not even print to the tape! The two counter values of a simulated 2-counter machine are encoded by the $(x,y)$ position of the tape head on the 2D tape. Testing for zero amounts to detecting one of the axes. It is well-known that 2-counter machines are universal~\cite{Minsky1967}. However, using known algorithms, 2-counter machines suffer from a doubly-exponential slowdown when simulating Turing machines~\cite{Schroeppel1972}, and so the 2-state, 3-symbol machine of Kleine-B\"uning and Ottmann also suffers from a doubly-exponential slowdown when simulating Turing machines. 
We give a brief overview of this machine's computation.

The 2D tape uses only the upper-right quadrant of the plane and so each tape cell may be indexed by a coordinate of the form $(x,y) \in  \mathbb{N} \times \mathbb{N}$. 
The quadrant is filled using 4, infinitely repeated, finite square blocks (of tape symbols) which we will call A, B, C, and D. 
The infinite pattern on the 2D tape given by the arrangement of these blocks is ultimately periodic in both the $x$ and $y$ directions. 
Each block is of size $O(r^2)$ where $r$ is the number of instructions in the 2-counter machine being simulated. 
The block at the origin of the quadrant is of type A. Types B and C are repeated along along the $x$-axis and $y$-axis, respectively, and the remainder of the quadrant is tiled by blocks of type D. 
Each block encodes the entire program of the 2-counter machine being simulated. 
The current counter machine instruction being simulated is given by the position of the tape head within a block. 
If the counters have values $x_1$ and $y_1$ respectively, then the tape head will be in the $x_1^{\textrm{th}}$ block from the $y$-axis and the $y_1^{\textrm{th}}$ block from the $x$-axis. 
The blocks contain specially defined paths that the tape head follows to (a) arrive at the next counter machine instruction and (b) move to one of the adjacent blocks if a change in the value of a counter is being simulated. 
A, B and C blocks lie along the axes and so are used to simulate any instruction where one or more counters have value zero, and in particular they contain special paths that simulate a positive test for zero.

Kleine-B\"uning and Ottmann adapt their technique to give a non-printing 1-state, 7-symbol universal machine with a 3D tape.
Only 2 planes in the third dimension are used, giving tape cells that are indexed by coordinates $(x, y, z)$, where $x,y\in\mathbb{N}$ and $z\in\{0,1\}$. 
The pattern defined by the symbols on each of the infinite 2D planes given by $(x,y,0)$ and $(x,y,1)$ is ultimately periodic in both the $x$ and $y$ directions. 
The technique used to simulate 2-counter machines by the 1-state, 7-symbol machine is, in essence, the same as the technique use by the 2-state, 3-symbol machine. 
The 2D machine uses 2 states to remember which path it is following when two different paths cross (the tape head follows paths that encode instructions of the counter machine being simulated). 
With the introduction of a third dimension it is no longer necessary for paths to cross and so it is possible to give a universal Turing machine with only~1 state. 
Finally, we note that an immediate corollary of this machine's design is the existence of a non-halting universal machine with only 6 symbols, as the only purpose of one of the 7 symbols is to provide an undefined transition rule for halting.

It is a fairly straightforward matter to show that for each Turing machine with a single, ultimately periodic, 2D tape and no print instructions there is a 2-counter machine that simulates it in linear time. 
It immediately follows that improvement on the doubly-exponential time overhead when simulating Turing machines with such non-printing 2D machines is not possible unless such an improvement is also possible for 2-counter machines. 
Thus, it could be interesting to see if the simulation time overhead for such machines can be reduced to singly-exponential when a slightly more complicated background is permitted on the tape.   

Wagner~\cite{Wagner1973} shows that the halting problem for Turing machines with a single $k$D tape ($k \in \mathbb{N}$), 2 symbols and 2 states is decidable\footnote{Machines using Wagner's definition end their computation with a simple loop: repeatedly executing a special transition rule that does not change the configuration. This is equivalent to executing a halting transition rule.}.  
Specifically, he shows that if such machines halt then they do so in space $O(n)$, where~$n$ is input length.
It is not difficult to give relevant decidability results (such as predicating looping or halting) for machines with a single 1D tape and non-printing instructions, even when an ultimately periodic background is permitted. 
Regarding $k$D machines, it can be shown for some classes of these machines that only weaker forms of universality are possible.  
For the case of $k$D non-printing machines, it is not difficult to give relevant decidability results when the initial tape contains only a finite number of non-blank symbols.  
Herman has shown that the halting problem is decidable for 1-state $k$D printing machines when all but a finite number of tape cells are blank at the start of each computation~\cite{Hermann1968}.

Though lacking in formal rigour, a comparison between the three 2D machines we discussed in this section poses some interesting questions about the possible trade-offs for different 2D models. For example, out of the three machines the 2-state, 3-symbol weak machine has the smallest state-symbol product, is the only non-writing machine, and the only machine that can halt. The 2-state, 4-symbol machine of Priese is the only machine of the three that does not use a periodic (weak) encoding, and the 4-state, 2-symbol machine of Gajardo et al.~(Langton's ant) is the only machine of the three that simulates Turing machines in polynomial time.  
The best we can hope for with non-printing 2D machines is a singly exponential time overhead, but achieving even this bound would seem to be very tricky. It is interesting to note that the only non-weak 2D machine of the three, that of Priese, has an exponential time overhead when simulating Turing machines. 
This is not the case for the smallest non-weak 1D machines. 
It begs the question, is there a non-weak 2D machine with the same number of states and symbols as Priese's machine that is universal with a polynomial time overhead? 

\subsection{Termination of a computation}
As we hope has been made clear so far, it is vitally important to clearly specify the computational model one is using when trying to find small universal programs or give lower bounds on universal program size. In the absence of a clear model description and matching lower bounds, one can never claim to have found the ``smallest'' universal program. 
Throughout this work we have described results on upper bounds and lower bounds on universal program size and we have described how both change when the model definition changes. In this section we focus on one such issue: computation termination.

A number of authors have given universal Turing machines where successful computations do not end in a halt state. Many of the machines given in Table~\ref{tab:non-standard} are non-halting. What about the problem of proving relevant non-universality results for these models? Such non-universality results are not achievable by proving the halting problem decidable. Before we attempt such an endeavour we must agree on a clear definition of universal Turing machine. For example, instead of specifying the end of a computation by a single halting (or terminal) configuration, a computation could end with a specific sequence of configurations. We refer to this as a terminal configuration sequence. The output of the simulated Turing machine is retrieved by applying a recursive decoding function to the entire computation (also a configuration sequence). There are many ways to define terminal configuration sequence, some examples are:
\begin{itemize}
\item a configuration sequence that goes through a specified sequence of states,
\item a configuration sequence that contains two identical configurations,
\item a configuration that contains a specific subword.
\end{itemize}
Given a definition of a terminal configuration sequence we may prove that the terminal sequence problem (will a machine execute a terminal configuration sequence) is decidable. This gives non-universal lower bounds that are relevant to universal machines that end their computation with such a sequence. However, this result may not hold as a proof of non-universality if we subsequently alter our definition of terminal configuration sequence. One more general approach is to prove that the terminal sequence problem for all possible terminal sequences, of a machine or set of machines, is decidable. In any case, it is important to specify these details when giving upper and lower bounds on program size.

\section{Busy beavers}
Besides small universal Turing machines, one finds small, yet complicated, programs in the busy beaver literature. The term busy beaver was introduced by Rado~\cite{Rado1962} who put forward a game where the goal for a given $k\in\mathbb{N}$ is to find, out of all the $k$-state, 2-symbol Turing machines, the machine that prints the most 1s and then halts when started on a blank tape. The busy beaver function $\Sigma: \mathbb{N} \rightarrow \mathbb{N}$ is then defined by letting $\Sigma(k)$ be the maximum number of 1's printed by any halting $k$-state, 2-symbol Turing machine. Busy beavers essentially adhere to the standard Turing machine model described in previous sections (one tape, one head, usual blank symbol, deterministic). It is known that $\Sigma(1) = 1$ (trivial), $\Sigma(2) = 4$~\cite{Rado1962}, $\Sigma(3) = 6$~\cite{LinRado1965}, and $\Sigma(4) = 13$~\cite{Brady1983}. However for $5$ states or more the best we currently have are lower bounds. 
For example, Michel~\cite{Michel2010} cites $\Sigma(5) \geq 4098$ to Marxen and Buntrock~\cite{MarxenBuntrock1990}, and $\Sigma(6) \geq 3.5 \times 10^{18267}$ to Pavel Kropitz. $S(k)$, the step-counting analogue of $\Sigma(k)$, is also considered. In fact, both~$\Sigma$ and~$S$ grow faster than any computable function~\cite{Rado1962}. Green~\cite{Green1964} has given a lower bound on the growth of the function $\Sigma$.

The busy beaver problem has been generalised to machines with $\ell \geq 2$ symbols~\cite{Brady1988}, where $\Sigma(k,\ell)$ is the largest number of non-zeros written by any $k$-state, $\ell$-symbol Turing machine. It has been shown~\cite{Brady1988,LafittePapazian2007} that $\Sigma(2,3) = 9$. Terry Ligocki and Shawn Ligocki have shown that $\Sigma(2,4) \geq 2,050 $ and $\Sigma(3,3) \geq 374,676,383$, and have given lower bounds on a number of other state-symbol pairs. See Michel's survey~\cite{Michel2010} for more results.

Although finding busy beavers is somewhat orthogonal to the goal of finding small universal Turing machines, there are potential connections between the two fields. On the one hand, when designing small universal programs one often has to reuse instructions in many different contexts, something which busy beavers might also do, so perhaps small instruction sets from one field might be useful for the other. On the other hand, proving lower bounds on universal program size, and upper bounds on values for the busy beaver function, both involve hefty case analyses so once again techniques developed in one field could potentially be useful for the other. In particular, the search for busy beavers has produced small programs with very complicated behaviour, which lend weight to the idea that proving non-universality of such program classes might be difficult.

\section{Further work}
There are many avenues for further work, here we highlight a few examples.

Applying computational complexity theory to the area of small universal Turing machines allows us to ask a number of questions that are more subtle than the usual questions about program size. 
As we move towards the origin in Figure~\ref{fig:states-sym-jan07}, the universal machines have larger (but polynomial) time overheads. 
Can the time overheads in Figure~\ref{fig:states-sym-jan07} be further improved (lowered)?  
Can we prove lower bounds on the simulation time of machines with a given state-symbol pair?  
Proving non-trivial simulation time lower bounds seems like a difficult problem. Such results could be used to prove that there is a polynomial trade-off between simulation time and universal program size. 

As we move away from the origin, the non-universal machines seem to have more power. For example Kudlek's classification of 2-state, 2-symbol machines shows that the sets accepted by these machines are regular, with the exception of one context free language ($a^{n} b^{n}$). Can we hope to fully characterise the sets accepted by non-universal machines (e.g.~in terms of complexity or automata theoretic classes) with given state-symbol pairs or other program restrictions?

When discussing the complexity of small machines the issue of encodings becomes very important. For example, when proving that the prediction problem for a small machine is P-complete~\cite{GHR1995x}, the relevant encodings should be in logspace, and this is the case for all of the polynomial time machines in Figure~\ref{fig:states-sym-jan07}.

Of course there are many models of computation that we have not mentioned where researchers have focused on finding small universal programs. Post's~\cite{Post1943} tag systems are an interesting example. Minsky~\cite{Minsky1960A,Minsky1961} showed that tag systems are universal with deletion number 6. Cocke and Minsky lowered the deletion number to 2, by showing that 2-tag systems were universal. They used productions (appendants) of length at most 4. Wang~\cite{Wan1963p2} further lowered the production length to 3. Recently, De Mol~\cite{Dem2007c} has given a lower bound by showing that the reachability (and thus halting) problems are decidable for 2-tag systems with 2 symbols; a problem which Post claimed~\cite{Pos1965m} to have solved but never published. It would be interesting to find the smallest universal tag systems in terms of number of symbols, deletion length, and production length.

The space between the non-universal curve and the smallest non-weakly universal machines in Figure~\ref{fig:states-sym-jan07} contains some complicated beasts. These lend weight to the feeling that finding new lower bounds on universal program size is tricky. Most noteworthy are the weakly and semi-weakly universal machines discussed earlier. Table~\ref{tab:non-standard} highlights that the existence of general models that provably have less states and symbols than the standard universal machines can have (for example the machines with (states, symbols, dimensions, tapes) of (2,3,2,1)~\cite{KleineOttmann1977}, (1,7,3,1)~\cite{KleineOttmann1977}, and (1,2,1,4)~\cite{Hooper1969}). Also of importance are the busy beavers~\cite{Michel2010} and small machines of Margenstern~\cite{Margenstern1998,Mar2000p}, Baiocchi~\cite{Baiocchi1998}, and Michel~\cite{Michel1993,Mic2004p} that live in this region and simulate iterations of the $3x+1$ problem and other Collatz-like functions. So it seems that there are plenty of animals yet to be tamed.

\end{document}